\documentclass[usenatbib,onecolumn]{mn2e}

\newcommand{\gbf}       {\mbox{\boldmath$g$}}
\newcommand{\wbf}       {\mbox{\boldmath$w$}}
\newcommand{\Bbf}       {\mbox{\boldmath$B$}}
\newcommand{\kbf}       {\mbox{\boldmath$k$}}

\newcommand{\fbf}       {\mbox{\boldmath$f$}}
\newcommand{\rbf}       {\mbox{\boldmath$r$}}

\newcommand{\vbf}       {\mbox{\boldmath$v$}}

\newcommand{\omegabf}       {\mbox{\boldmath$\omega$}}

\newcommand{\Omegabf}       {\mbox{\boldmath$\Omega$}}

\def\gcc{\hbox{\rm\hskip.35em  g cm}$^{-3}$}
\def\rads{\hbox{\rm\hskip.35em  rad s}$^{-1}$}
\def\cms{\hbox{\rm\hskip.35em  cm s}$^{-1}$}

\def\eg{{\it e.g.}}
\def\lap{\hbox{${_{\displaystyle<}\atop^{\displaystyle\sim}}$}}
\def\gap{\hbox{${_{\displaystyle>}\atop^{\displaystyle\sim}}$}}

\usepackage{graphicx}
\usepackage{subfigure}

\title[Instability of Superfluid Flow]
{Instability of Superfluid Flow in the Neutron Star Core}
\author[B. Link]{B. Link\thanks{E-mail:
link@physics.montana.edu;}\\
Department of Physics, Montana State University, Bozeman, MT
59717, USA}
\begin{document}


\pagerange{\pageref{firstpage}--\pageref{lastpage}} \pubyear{2012}

\maketitle

\label{firstpage}

\begin{abstract}
Pinning of superfluid vortices to magnetic flux tubes in the outer
core of a neutron star supports a velocity difference of $\sim 10^5$
\cms\ between the neutron superfluid and the proton-electron fluid as the 
star spins down. Under the
Magnus force that arises on the vortex array, vortices undergo {\em
vortex creep} through thermal activation or quantum tunneling.  We
examine the hydrodynamic stability of this situation. Vortex creep
introduces two low-frequency modes, one of which is unstable above a
critical wavenumber for any non-zero flow velocity of the neutron
superfluid with respect to the charged fluid. For typical pinning
parameters of the outer core, the superfluid flow is unstable over
wavelengths $\lambda\lap 10$ m and over timescales of $\sim
(\lambda/\mbox{1 m})^{1/2}$ yr down to $\sim 1$ d. The vortex lattice
could degenerate into a tangle, and the superfluid flow would become
turbulent. We suggest that superfluid turbulence could be responsible
for the red timing noise seen in many neutron stars, and find a
predicted spectrum that is generally consistent with observations.
\end{abstract}


\section{Introduction}
\maketitle


The dynamical role of hydrodynamic instabilities in the liquid core of
a neutron star is of considerable interest in understanding a
variety of phenomena that includes spin glitches, stochastic spin
variations (``timing noise'') and thermal evolution, as well as
possible precession and r-modes. One instability that could arise is
the ``Glaberson-Donnelly'' counterflow instability observed in
superfluid liquid helium, wherein flow of the fluid's normal component
along the vortices that thread a rotating superfluid drives the system
turbulent \citep{Glaberson_etal74}. Early in a neutron star's
evolution, when the temperature of the core is comparable to the
condensation temperature of the neutrons, differential rotation in the
core, resulting from a glitch or possibly causing it, drives an Ekman
flow along the rotation axis that can excite the Glaberson-Donnelly
instability
\citep{peralta_etal05,peralta_etal06}. \citet{peralta_etal06} and
\citet{mp07} have shown that transitions between
laminar flow and fully-developed turbulence could drive spin glitches.
Unstable shear layers in the outer core may also play a role in
glitches \citep{pm09}.

The protons of the outer core are predicted to form a type II
superconductor \citep{migdal59}. When the protons condensed early in
the star's life, the expulsion time for magnetic flux was far greater
than the nucleation time of magnetic flux tubes, and the magnetic flux
became confined to flux tubes \citep{bpp69}. The magnetic field inside
a flux tube is comparable to the lower critical field for
superconductivity, $B_c\sim 10^{15}$ G. Entrainment of the proton and
neutron mass currents causes magnetization of the neutron vortices
with a field in the vortex core comparable to $B_c$ \citep{als84}. If
the vortices are free to move, momentum exchange between the charged
fluid and the neutron superfluid through the scattering of electrons
with the magnetized vortex cores quickly damps velocity differences
between the two fluids. The strong magnetization of the vortex cores,
however, will pin the vortices to the flux tubes with a pinning energy
of $\gap 100$ MeV per intersection (see
\citealt{jones91,mendell91a,chau_etal92,rzc98}, and the estimates of
\S 2 below). Pinning plays a crucial role in the dynamics of the core
fluids; to the extent that pinning is perfect, the motions of the two
fluids are largely independent, apart from modifications to the
average mass currents by the entrainment effect.

\citet{sac08} and \citet{gaj08a,gaj08b} have studied a variant of the
Glaberson-Donnelly instability for the core mixture of superfluid
neutrons, superconducting protons, and electrons. The proton-electron
liquid acts analogously to the normal component in liquid helium. They
found that inertial waves of the neutron superfluid are subject to the
Glaberson-Donnelly instability if there is a relative flow between the
proton-electron fluid and the neutron fluid directed along the
rotation axis, whether pinning is negligible \citep{sac08} or perfect
\citep{gaj08a,gaj08b}. Such a relative flow could result in a
precessing neutron star in which pinning of vortices to flux tubes is
effective. \citet{vl08} have shown, however, that the enormous
magnetic stress in the charged fluid suppresses the instability for
relative flow speeds less than the hydromagnetic wave speed; the
instability is unlikely to occur, even in a precessing neutron star.

\citet{acp04} have shown that relative motion between the neutron
fluid and the charged fluid can be susceptible to a two-stream
instability, though pinning was not considered. This instability is
similar to the Kelvin-Helmholtz instability for relative flow of two
fluids separated by an interface, but occurs when the fluids coexist
in the same volume.
\citet{ga09} have identified a 
instability that can occur in the neutron-proton mixture of the core
associated with inertial r-modes, assuming perfect pinning of vortices
to flux tubes. The instability occurs if the relative angular velocity
difference between the proton and neutron liquids everywhere exceeds a
critical value. \citet{ga09} propose that this instability triggers
glitches, though it has not been explained how the the critical
angular velocity difference for instability reaches its critical value
nearly everywhere in the star at once, or how a glitch is subsequently
triggered.

From the standpoint of building realistic rotational models of neutron
stars, a crucial question is if laminar flow in the star is generally
stable or unstable. As a first step towards answering this question,
we studied the stability of the vortex lattice, pinned to the nuclei
of the inner crust, in the presence of the superfluid flow sustained
by pinning as the star spins down
\citep{link12a}. For the situation of {\em
imperfect pinning} that would arise as a result of, for example, 
vortex motion through thermal activation or quantum tunneling, two
low-frequency modes appear in the system, one of which is unstable
over length scales of less than $\sim 10$ m, and over timescales
possibly as fast as $\sim 100$ d. We suggested that the instability
could lead to turbulence similar to grid turbulence seen in liquid
helium \citep{smith_etal93}.  In this paper, we extend the analysis to
the neutron star core and find an instability that is analogous to
that found in crust. As the star spins down, the Magnus force on the
pinned vortices exerts stresses on the neutron and charged fluids that
drives the flow unstable, possibly over timescales as short as days. 

In \S 2, we estimate the angular velocity difference between the
neutron fluid and the charged fluid that can be sustained by pinning
in a spinning-down neutron star. In \S 3, we examine the stability of
this state of differential rotation, and show that it is unstable. In
\S 4, we estimate the spectrum of stochastic torque that would arise
if the fluid becomes turbulent, and compare to measured power spectra
in radio pulsars and magnetars. Our chief conclusion is that the
superfluid in a spinning-down neutron star is generally turbulent
everywhere that there is significant pinning, hence, throughout the
outer core and inner crust. 

\section{Pinning of vortices to flux tubes}

We begin by establishing the approximate density range of type II
superconductivity, where the magnetic field will be confined to flux
tubes and the pinning of vortices to flux tubes will be important.
The proton fluid will be a type II superconductor when the proton
coherence length $\xi_p$ and the London length $\Lambda_*$ satisfy
\begin{equation}
\xi_p<\sqrt{2}\Lambda_*.
\end{equation}
The London length is \citep{als84}
\begin{equation}
\Lambda_*=30\, 
\left[
\frac{m_p^*}{m_p}x_p^{-1}\rho_{14}^{-1}
\right]^{1/2}\mbox{ fm}, 
\label{london}
\end{equation}
where $m_p$ is the bare neutron mass and $m_p^*$ is its effective mass
in the medium, $x_p\equiv \rho_p/\rho_n$, $\rho_p$
and $\rho_n$ are the mass densities of protons and neutrons, and
$\rho_{14}$ is the total mass density in units of $10^{14}$ \gcc. 
In the outer core, $m_p^*/m_p\simeq 1/2$ \citep{sjoberg,ch06}. 
 
The proton coherence length $\xi_p$ is \citep{mendell91a}
\begin{equation}
\xi_p=16\,
x_p^{1/3}\rho_{14}^{1/3}\,\frac{m_p}{m_p^*}\,\Delta_p(\mbox{MeV})^{-1}
\mbox{ fm}, 
\label{xip}
\end{equation}
where $\Delta_p$ is the proton pairing gap. The core begins at a
density of $\rho_{14}\simeq 1.5$ \citep{rpw83}, and the proton gap $\Delta_p$
has a typical value of $\sim 1$ MeV around nuclear density 
$\rho_{14}=2.8$ \citep{relgap}. Combining
eqs. (\ref{london}) and (\ref{xip}) gives
\begin{equation}
\frac{\xi_p}{\sqrt{2}\Lambda_*}\simeq 0.3\,
\left(\frac{m_p^*/m_p}{0.5}\right)^{-3/2}\left(\frac{x_p}{0.05}\right)^{5/6}
\left(\frac{\rho_{14}}{4}\right)^{5/6}\,
\Delta_p(\mbox{MeV})^{-1}.
\end{equation}
The protons thus form a type II superconductor in the outer core.
Above nuclear density the proton gap begins to fall \citep{relgap},
and a transition to a type I superconductor occurs at several times
nuclear density. Polarization effects may lower $\Delta_p$ by a
factor of 2-3 \citep{schulze_etal96}, restricting the type II region
to somewhat lower density.

Vortices and flux tubes interact with one another and pin as a result
of the interaction between the neutron and proton
condensates. For neutrons flowing at velocity $\vbf_n$ and protons at
velocity $\vbf_p$, the mass currents take the form \citep{als84}
\begin{equation}
\gbf_p=\rho_{pp}\vbf_p+\rho_{pn}\vbf_n,
\qquad
\gbf_n=\rho_{nn}\vbf_p+\rho_{np}\vbf_n,
\end{equation}
The mass current of each species is generally in a direction different
than the velocity of either species. This {\em entrainment effect} is
fundamentally non-dissipative.

For a proton mass density $\rho_p$ and neutron mass density $\rho_n$,
the coefficients for the mass currents are given by
\begin{equation}
\rho_{pp}=\rho_p\left(\frac{m_p}{m_p^*}\right) \qquad
\rho_{nn}=\rho_n\left(\frac{m_n}{m_n^*}\right) \qquad
\rho_{np}=\rho_{pn}=\rho_p\left(\frac{\delta m_p^*}{m_p}\right)
=\rho_n\left(\frac{\delta m_n^*}{m_n}\right), 
\end{equation}
where $m_n^*$ is the effective mass of the 
neutron; $\delta m_p^*\equiv m_p^*-m_p$ and $\delta m_n^*\equiv
m_n^*-m_n$ are the contributions to the effective masses due to
interactions in the medium. The neutron and proton mass densities, and
the total density, neglecting the electron mass, are
\begin{equation}
\rho_p=\rho_{pp}+\rho_{pn}
\qquad
\rho_n=\rho_{nn}+\rho_{np}
\qquad
\rho=\rho_n+\rho_p.
\end{equation}
In the outer core, $\delta m_p^*/m_p\simeq -1/2$ 
and $m_n^*/m_n\simeq 1$ \citep{sjoberg,ch06}, and $\rho_{pn}$ is
negative. 

\citet{rzc98}, using Ginzburg-Landau
theory and the results of \citet{als84}, have calculated the 
energy per unit length $L$ of a superimposed flux tube and
vortex line, minus the energy for infinite separation: 
\begin{equation}
\frac{E_0}{L}\simeq \frac{\pi}{8}\left(\frac{\Phi_0}{\pi\Lambda_*^2}\right)^2
\Lambda_*^2\,\frac{m_p}{m_n}\frac{\rho_{pn}}{\rho_{pp}}
\ln\left(\frac{\Lambda_*}{\xi}\right)
=\frac{\pi}{8}B_v B_\Phi\Lambda_*^2
\ln\left(\frac{\Lambda_*}{\xi_n}\right), 
\label{EL}
\end{equation}
where $\Phi_0\equiv hc/2e$ is the flux quantum,
$B_\Phi\equiv\Phi_0/\pi\Lambda_*^2$ is the characteristic magnetic
field in a flux tube core,
$B_v\equiv(\Phi_0/\pi\Lambda_*^2)(m_p\rho_{pn}/m_n\rho_{pp})$ is the
field in the vortex core, and
$\xi_n$ is the neutron coherence length.  Since $\rho_{pn}$ is
negative, $E/L$ is negative if the vorticity is parallel to $\Bbf_v$,
and positive if anti-parallel.  The interaction is primarily
magnetic. Both the flux tube and the vortex line have their magnetic
fields screened over the London length $\Lambda_*$.

If the angle between the vortex line and the flux tube is $\theta$,
the overlap length is $l\simeq 2\Lambda_*/\sin\theta$. 
The interaction energy per vortex-flux tube junction is
approximately 
\begin{equation}
E_p(\theta)\simeq
l\,\frac{E_0}{L}=
\frac{\pi}{8}\,\Bbf_v\cdot\Bbf_\Phi\,(\Lambda_*^2l)
\ln\left(\frac{\Lambda_*}{\xi_n}\right) =
\frac{\pi}{4}\,\left(\frac{\Phi_0}{\pi\Lambda_*^2}\right)^2
\Lambda_*^3\,\frac{\delta m_p^*}{m_p}\,
\ln\left(\frac{\Lambda_*}{\xi_n}\right)\, \cot\theta
\end{equation}
This equation provides only a rough estimate for arbitrary $\theta$, since
eq. (\ref{EL}) does not account for modifications of vortex and flux
tube structure for $\theta\ne0$.  The pinning energy is approximately
the magnetic energy density in the overlap region, times the overlap
volume $\Lambda_*^2 l$. Similar estimates have been obtained by 
\citet{jones91}, \citet{mendell91a}, and \citet{chau_etal92}.

The magnetic field is expected to be highly tangled \citep{rzc98}, and so
$\theta$ could take a broad range of values. For an average radius of 
curvature of flux tubes of $R_\Phi$, the maximum overlap length is 
\begin{equation}
l_{\rm max}=2R_\Phi \cos^{-1}(1-\Lambda_*/R_\Phi)\equiv
\frac{2\Lambda_*}{\sin\theta_{\rm min}}, 
\end{equation}
where $\theta_{\rm min}$ is the minimum angle at which a curved flux
tube intersects a straight vortex. 
For $R_\Phi>>\Lambda_*$, this equation gives 
$\theta_{\rm min}\simeq (2\Lambda_*/R_\Phi)^{1/2}$. 
The interaction
is repulsive for $\pi/2<\theta<\pi$, 
but the magnitude of the pinning interaction is the same
for both $\theta$ and $\pi-\theta$. Vortices will be immobilized by
flux tubes in much the same way regardless of the sign of the
interaction. We average $\vert E_p(\theta)\vert$ over
$\theta$ to obtain the average pinning energy for $R_\Phi>>\Lambda_*$
\begin{equation}
\langle E_p\rangle 
=\frac{1}{(\pi-2\theta_{\rm min})}\int_{\theta_{\rm
min}}^{\pi-\theta_{\rm min}}d\theta\, \vert E_p(\theta)\vert
\simeq \frac{2}{\pi}\, E_p(\cot\theta=1)\,	
\ln\sqrt{\frac{R_\Phi}{2\Lambda_*}}. 
\label{eave}
\end{equation}
Unless $R_\Phi$ is much larger than $\sim 10^3\Lambda_*$, the
logarithmic factor is not significantly different than one, so we
henceforth set this factor to unity. If $R_\Phi$ is much larger,
however, it is probably a better approximation to set $\theta$ equal
to the inclination angle between the star's magnetic moment and the
rotation axis. In either case eq. (\ref{eave}), with the logarithmic
factor set to unity, should be a good estimate. 

The pinning energy per vortex-flux tube junction is, 
taking a typical value $\ln(\Lambda_*/\xi_n)=0.5$, 
\begin{equation}
\langle E_p\rangle\simeq 10^2\, 
\left(\frac{m_p^*/m_p}{0.5}\right)^{-1/2}
\left(\frac{\vert\delta m_p^*\vert/m_p}{0.5}\right)
\left(\frac{x_p}{0.05}\right)^{1/2}
\left(\frac{\rho_{14}}{4}\right)^{1/2}
\mbox{ MeV}, 
\end{equation}
The pinning force is $F_p\sim \langle E_p\rangle/\Lambda_*$, 
typically $\sim 1$ MeV fm$^{-1}$. 

For a single vortex immersed in a tangle of flux tubes, the average
length between intersections will equal the average distance between
flux tubes, $l_\Phi=n_\Phi^{-1/2}$, where $n_\Phi=B/\Phi_0$ is the
areal density of flux tubes and $B$ is the average magnetic field
strength. The pinning force per unit length is $f_p=F_p/l_\Phi$.  For
a velocity difference $w$ between the neutron superfluid and the
pinned vortex, the fluid flow exerts a Magnus force per unit length
$\rho_n\kappa w$, where $\kappa\equiv h/2m_n$ is the vorticity
quantum. The critical velocity difference $w_c$ that can be sustained
is given by
\begin{equation}
f_p=\frac{\langle E_p\rangle}{\Lambda_* l_\Phi}=\rho_n\kappa w_c
\end{equation}
so that
\begin{equation}
w_c\sim 10^5\, 
\left(\frac{x_p}{0.05}\right)
\left(\frac{m_p^*/m_p}{0.5}\right)^{-1}
\left(\frac{\vert\delta
m_p^*\vert/m_p}{0.5}\right)\,B_{12}^{1/2}\mbox{ \cms}, 
\label{wc}
\end{equation}
where $B_{12}$ is the strength of the field in units of $10^{12}$ G. 
The critical velocity is not a strong function of density. 

As a star spins down, pinning prevents the neutron superfluid from
corotating with the charged fluid, and a velocity difference $w$
builds. If $w\simeq 0$ at some time, spin down of the crust will cause
the critical velocity to be reached in a time $\Delta t$ that is short
compared to the star's age:
\begin{equation}
\Delta t\simeq \frac{w_c}{R\,2\pi\vert\dot\nu\vert}=
30\, 
\left(\frac{w_c}{10^5\mbox{ \cms}}\right)
\left(\frac{t_{\rm age}}{10^4\mbox{ yr}}\right)
\left(\frac{\nu}{10\mbox{ Hz}}\right)^{-1}\mbox{ yr}, 
\end{equation}
where $\nu$ is the spin frequency of the crust, $t_{\rm
age}\equiv\nu/2\dot\nu$ is the spin-down age, and $R$ is the stellar
radius. 
Hence, the angular velocity
difference between the superfluid and the charged components will be
$w_c/R\sim 10^{-1}$ \rads\ throughout the star's life (except,
for example, right after a glitch). 
We now examine the stability of this state of differential rotation. 

\section{Stability analysis} 

The problem of the coupled dynamics of the neutron and proton fluids
can be treated with hydrodynamics for length scales that are large
compared to the inter-vortex spacing $l_v$, restricting the treatment
to wavenumbers $kl_v<<1$. In a single-component superfluid, the vortex
lattice supports Tkachenko modes of speed $c_T=(\hbar\Omega/4m)^{1/2}$
\citep{tk2a,tk2}, where $\Omega$ is the spin rate of the
superfluid and $m$ is the mass of the fundamental boson, twice the
neutron mass for a neutron superfluid. (For studies of Tkachenko modes
in neutron stars and the effects of dissipation on these modes, see
\citealt{ns08} and \citealt{haskell11}).  For a typical neutron star,
$c_T\simeq 10^{-1}\, (\Omega/100\mbox{ \rads})^{1/2}$ \cms. The
degrees of freedom of the vortex lattice can be ignored to a very good
approximation when the speed of the background flow with respect to
pinned vortices is much larger than $c_T$ \citep{link12a}, as is the
case for vortex pinning to flux tubes. The areal density of vortices
is $l_v^{-2}=2m\Omega/h$ for a uniform vortex lattice; hence, the
requirement that $kl_v<<1$ is equivalent to $c_Tk<<\Omega$.

In the laboratory frame, the acceleration equations for neutrons flowing at
velocity $\vbf_n$ and protons plus electrons flowing at velocity 
$\vbf_p$, are \citep{prix04,ac06,gaj08a,vl08,gas11} 
\begin{equation}
(\partial_t +\vbf_n\cdot\nabla)(\vbf_n-\epsilon_n\wbf_{np})
-\epsilon_n w_{np}^i\nabla v_i^n
=
-\nabla\mu_n +\fbf/\rho_n
\end{equation}
\begin{equation}
(\partial_t +\vbf_p\cdot\nabla)(\vbf_p+\epsilon_p\wbf_{np})
+\epsilon_p w_{np}^i\nabla v_i^p =
-\nabla\mu_p-\fbf/\rho_p
+\nu_e\,\nabla^2\vbf_p+(1/4\pi\rho_p)\,\Bbf_{\rm
eff}\cdot\nabla\Bbf_{\rm eff},
\end{equation}
where $\wbf_{np}\equiv \vbf_n-\vbf_p$, $i$ is a coordinate index and
is summed, $\mu_n$ and $\mu_p$ are the neutron and proton chemical
potentials, $\rho_n$ and $\rho_p$ are the neutron and proton mass
densities (neglecting the electron mass), $\fbf/\rho_n$ is the force
per unit volume exerted on the neutron fluid by the proton-electron
fluid, $\nu_e$ is the electron kinematic shear viscosity, and $B_{\rm
eff}\equiv\sqrt{BB_c}$, where $B$ is the average field strength and
$B_c\simeq 10^{15}$ G is the value of the lower critical
field. Electron-electron scattering gives the dominant contribution to
the viscosity, as all other scattering processes are strongly
suppressed when both the neutrons and protons are superfluid. The
terms proportional to $\epsilon_n$ and $\epsilon_p$ account for
entrainment between neutrons and protons, and are related by
$\rho_n\epsilon_n=\rho_p\epsilon_p$. We take the entrainment
coefficients to be constants. The small effects of electron inertia
and the London current, both neglected here, have been considered by
\citet{gas11}. 

For the highly-conductive proton-electron fluid, the magnetic field is
frozen to the proton-electron fluid. The field obeys the 
induction equation
\begin{equation}
\partial_t\Bbf+\nabla\times(\Bbf\times\vbf_p)=0.
\end{equation}

The force per unit volume on the neutron fluid, $\fbf/\rho_n$, is
equal to the Magnus force per unit volume on the vortex array, that
is,
\begin{equation}
\fbf/\rho_n=\omegabf_n\times(\vbf_n-\vbf_v)
\label{force}
\end{equation}
where $\vbf_v$ is the local vortex velocity and
$\omegabf_n\equiv\nabla\times\vbf_n$ is the local vorticity of the
neutron superfluid. For
perfect pinning of vortices against flux tubes, $\vbf_v=\vbf_p$. The
vortex velocity with respect to the proton-electron fluid has a
component along $\wbf_{np}\equiv\vbf_n-\vbf_p$, and a component
orthogonal to both $\wbf_{np}$ and 
$\omegabf_n$. The local vortex velocity can thus be written as
\begin{equation}
\vbf_v=\vbf_p+(1-\beta^\prime)\,\wbf_{np}+
\beta\,\wbf_{np}\times\hat{\omega}_n, 
\label{vv}
\end{equation}
where $\beta^\prime$ and $\beta$ are coefficients that determine the
vortex mobility, and we assume to be constant. Perfect pinning of
vortices to the charged fluid 
corresponds to $\beta^\prime=1$ and $\beta=0$.  The force, from
eq. (\ref{force}), is
\begin{equation}
\fbf/\rho_n=\beta^\prime\,\omegabf_n\times\wbf_{np}
+\beta\, \hat{\omega}_n\times(\omegabf_n\times\wbf_{np}).
\label{fmf}
\end{equation}
This force is the mutual friction force introduced in by \citet{hv56},
but appearing here in a different context. Imperfect pinning, that is,
``vortex creep'', corresponds to $\alpha\equiv1-\beta^\prime<<1$ and
$\beta<<1$.  We refer to $\alpha$ and $\beta$ as the 
``pinning coefficients''.  Perfect pinning corresponds to the limit
$\alpha=\beta=0$, while no pinning ($\fbf=0$) corresponds to
$\alpha=1$ and $\beta=0$.  Vortices move with a component along
$\wbf_{np}$, so that $0<\alpha\le 1$. The energy dissipation rate
per unit volume is determined by $\beta$, which must be positive to
give local entropy production. More general dissipative forces have
been considered by \citet{ac06}. 

Vortex creep could be a low-drag process, with $\beta<<\alpha$, or a
high-drag process, with $\beta>>\alpha$. In
much previous work on pinning, the high-drag limit has been implicitly
assumed through the following relationship between 
$\beta$ and $\beta^\prime$: 
\begin{equation}
\beta^\prime= 1-\alpha=\frac{{\cal R}^2}{1+{\cal R}^2}={\cal R}\,\beta, 
\label{drag}
\end{equation}
where ${\cal R}$ is a dimensionless drag coefficient. In this drag
description, imperfect pinning corresponds to ${\cal R}>>1$ so that
eq. (\ref{drag}) 
{\em requires} $\beta>>\alpha$. Eq. (\ref{drag}) is not true in
general, since the presence of non-dissipative forces between vortices
and the charged fluid can give $\beta<<\alpha$, a
regime of low drag \citep{link09}. As we show below, it is the
low-drag regime that is likely to be realized, with vortex creep being
unstable in this regime. A crucial feature of our analysis is that we
do not impose eq. (\ref{drag}) for imperfectly-pinned vortices. We
will assume in \S 4 that eq. (\ref{drag}) holds only for the tiny fraction of
vortex length ($\sim 10^{-6}$) that is unpinned at any instant. 

To examine the stability of the flow with imperfect pinning, we use a
local plane wave analysis in the frame rotating with the unperturbed
charged fluid at angular velocity $\Omegabf_0$, in which $\vbf_p=0$
and the unperturbed flow velocity arising from spin down of the
charged fluid is $\wbf_0=\wbf_{np,0}=w_0\hat{x}$. Restricting the
analysis to the regime $k R\gap 1$, where $R$ is the stellar radius,
the background flow can be treated as approximately uniform. Let the
rotation axis $\hat{z}$, $\kbf$, and $\wbf_0$ all be coplanar, with an
angle $\theta$ between $\kbf$ and the rotation axis. The angle between
$\hat{z}$ and $\Bbf$ is $\theta_B$. We restrict the analysis to the
quadrant $0\le\theta\le\pi/2$. The linearized equations of motion in the
rotating frame are:
\begin{equation}
(\partial_t+\wbf_0\cdot\nabla)(\delta
\vbf_n-\epsilon_n\delta\wbf_{np}) 
-\epsilon_n w_0\nabla (\hat{x}\cdot\delta\vbf_n)
-2\Omegabf_0\times\delta\vbf_n=-\nabla\delta\mu^\prime
+\delta\fbf/\rho_n
\label{naccel}
\end{equation}
\begin{equation}
\nabla\cdot\delta\vbf_n=0
\end{equation}
\begin{equation}
\partial_t(\vbf_p+\epsilon_p\delta\wbf_{np})
+\epsilon_p w_0\nabla (\hat{x}\cdot\delta\vbf_n)
+2\Omegabf_0\times\vbf_p=-\nabla\delta\mu_p^\prime
-\delta\fbf/\rho_p+\nu_e\nabla^2\vbf_p
+\frac{\Bbf_{{\rm eff},0}}{4\pi\rho_p}\cdot\nabla\delta\Bbf_{\rm eff}
\label{paccel}
\end{equation}
\begin{equation}
\delta\fbf/\rho_n=
(1-\alpha)\,\delta\{\omegabf_n\times\wbf_{np}\}
+\beta\,\delta\{\hat{\omega}_n\times(\omegabf_n\times\wbf_{np})\}.
\label{df}
\end{equation}
\begin{equation}
\nabla\cdot\vbf_p=0
\end{equation}
\begin{equation}
\partial_t\delta\Bbf+\nabla\times(\Bbf_0\times\vbf_p)=0
\label{induction}
\end{equation}
where $\delta$ denotes a perturbed quantity, and
$\delta\vbf_p=\vbf_p$. 
Here $\mu_{n,p}^\prime\equiv\mu_{n,p}-\rho_{n,p}(\Omegabf_0\times\rbf)^2/2$. 
The vorticity appearing in this equation is the total vorticity evaluated
in the laboratory frame, and is only slightly larger than $2\Omega_0$:
\begin{equation}
\frac{\omega_0-2\Omega_0}{2\Omega_0}\simeq \frac{w_0}{R\Omega_0}=
10^{-3}\,\left(\frac{w_0}{10^5\mbox{
\cms}}\right)\left(\frac{\Omega_0}{100\mbox{ rad
s$^{-1}$}}\right)^{-1}.
\end{equation}
We henceforth take $\omega_0=2\Omega_0$; we have confirmed that the
results presented below are essentially unchanged for
$\vert\omega_0-2\Omega_0\vert<<1$. 

For the shear perturbations we are considering,
$\kbf\cdot\delta\vbf_n=0$ and $\kbf\cdot\vbf_p=0$, that is, the
velocity perturbations in the directions $\hat{y}$ and $\hat{e}\equiv
-\cos\theta\,\hat{x}+\sin\theta\,\hat{z}$ are orthogonal to $\hat{k}$.
We Fourier transform $(\propto {\rm e}^{i{\mathbf k}\cdot{\mathbf
r}-i\sigma t})$ eqs. [\ref{naccel}]-[\ref{induction}] and project the
acceleration equations (\ref{naccel}) and (\ref{paccel}) onto
$\hat{y}$ and $\hat{e}$. Defining
$\sigma^\prime\equiv\sigma-kw_0\sin\theta$, $c\equiv\cos\theta$,
$s\equiv\sin\theta$, $v_B^2\equiv B_0B_c/4\pi\rho_p$,
$c_B\equiv\cos(\theta-\theta_B)$, and $x_p\equiv\rho_p/\rho_n$, we
obtain the system of equations:
\begin{equation}
 \left[ 
\begin{array}{cccc} 
{\cal A}-i\sigma^\prime-i\epsilon_n k w_0s & 
-2c\Omega_0\alpha &
-{\cal A}+i\epsilon_nk w_0 s & 
-2\Omega_0 c \beta^\prime \\
c(2\Omega_0\alpha-ik w_0 s\beta) & 
(\epsilon_n-1)\,i\sigma^\prime +2\Omega_0 c^2\beta-ikw_0s\beta^\prime & 
{\cal C} & 
-{\cal A}+i\epsilon_n kw_0s+2\Omega_0s^2\beta \\
-x_p^{-1}{\cal A} & 
-2x_p^{-1}\Omega_0 c \beta^\prime & 
{\cal B} &
2\Omega_0c(x_p^{-1}\beta^\prime-1) \\
x_p^{-1}{\cal C} & 
x_p^{-1}(-{\cal A}+2\Omega_0 s^2\beta) & 
2\Omega_0c-x_p^{-1}{\cal C} &
{\cal B}-2x_p^{-1}\Omega_0 s^2\beta \\
\end {array} 
\right]
\left[
\begin{array}{c}
\hat{y}\cdot\delta\vbf_n \\
\hat{e}\cdot\delta\vbf_n \\
\hat{y}\cdot\vbf_p \\
\hat{e}\cdot\vbf_p \\
\end{array}
\right]=0
\label{matrix}
\end{equation}
where
\begin{equation}
{\cal A}\equiv \epsilon_n\,i\sigma+2\Omega_0\beta-ikw_0s\beta^\prime,
\end{equation}
\begin{equation}
{\cal B}\equiv -i\sigma+iv_B^2c_B^2k^2/\sigma+\nu_ek^2 +x_p^{-1}{\cal
A},
\end{equation}
\begin{equation}
{\cal C}\equiv c(ikw_0s\beta+2\Omega_0\beta^\prime).
\end{equation}
(Recall that $\beta^\prime=1-\alpha$). The speed of hydromagnetic waves
is 
\begin{equation}
v_B=2\times 10^6\, \left(\frac{\rho_{14}}{4}\right)^{-1/2}
\left(\frac{x_p}{0.05}\right)^{-1/2}\,B_{0,12}^{1/2}\mbox{ \cms}, 
\label{vb}
\end{equation}
significantly greater than the speed $w_0$ of the background flow
(see eq. \ref{wc}). 

The dispersion relation that follows from eq. (\ref{matrix}) is an
impressively lengthy sixth-order polynomial in $\sigma$ in which all
coefficients from sixth order to zeroth order are non-zero; it is
easiest to explore various limits by working with eq. (\ref{matrix})
directly. We first confirm a previous result of \citet{vl08} for 
perfect pinning ($\alpha=\beta=0$), no background
flow ($w_0=0$), and no entrainment ($\epsilon_n=0$). In this case, the
polynomial factors into the form
\begin{equation}
\sigma^2(\sigma^4+C\sigma^3+(-A^2+2B+AC)\,\sigma^2+BC\sigma+B^2)=0. 
\label{dra}
\end{equation}
If we define 
\begin{equation}
f(\sigma)=\sigma^2+A\sigma+B=0
\end{equation}
and form the combination
\begin{equation}
\sigma^2(f(\sigma)^2+(C-2A)f(\sigma)\,\sigma)=0, 
\end{equation}
we obtain eq. (\ref{dra}), so two of the non-zero modes are given by the
the simpler quadratic expression. 
Reading $A$ and $B$ from eq. (\ref{matrix}) gives the dispersion
relation
\begin{equation}
\sigma^2+\left[-2\Omega_0\left(1-\frac{1}{x_p}\right)+i\nu_e
k^2\right]\sigma -\frac{4\Omega_0^2}{x_p}-v_B^2k^2=0, 
\label{drvl}
\end{equation}
as found by \citet{vl08}.\footnote{Our definition of $\sigma$ differs
from that of \citet{vl08} by a minus sign.}  For $\Omega_0=0$, the system has only damped
hydromagnetic waves that travel at speed $v_B$.

We will not present here an analysis of the full mode structure of the
system, but focus on two low-frequency modes that appear for imperfect
pinning. For small $\alpha$, $\beta$, and $\epsilon_n$, the two 
zero-frequency modes of eq. (\ref{dra}) become small, and we can
obtain these modes by working to second order in $\sigma$. 
Since $w_0$ is much smaller than $v_B$ (eqs. \ref{wc} and
\ref{vb}), we can further simplify the problem by taking the
limit $v_B\rightarrow\infty$ for finite $w_0$. We take this limit by
keeping only terms that multiply $v_B^4$, the highest order at
which $v_B$ appears. With these approximations, 
eq. (\ref{matrix}) gives
\[
(1-\epsilon_n)^2\sigma^2
+(2\epsilon_n^2 kw_0s
+\epsilon_n\{-2i\beta(1+c^2)-2\beta^\prime kw_0s\}
-2\alpha kw_0s+2i\beta (1+c^2))\sigma  
\]
\begin{equation}
\hspace*{2.cm}
-(\epsilon_n kw_0s)^2 
+\epsilon_n(2\beta^\prime (kw_0s)^2+
2i\beta\Omega_0kw_0 s (1+c^2))
-4(\alpha^2+\beta^2)\Omega_0^2 c^2 
+(\alpha kw_0s)^2-2i\alpha\beta kw_0 s = 0.
\label{dr}
\end{equation}
The electron viscosity does not appear at this level of approximation;
in the $v_B\rightarrow\infty$ limit, the flux tube array is
infinitely rigid, and vortex motion proceeds without producing shear
in the proton-electron fluid. The validity of this approximation is
confirmed below. 
The solutions to eq. (\ref{dr}) are
\begin{equation}
\sigma_\pm =\frac{1}{1-\epsilon_n}
\left(
(\alpha-\epsilon_n) kw_0\sin\theta-i\Omega_0(1+\cos^2\theta)\beta 
\pm \left(
4\Omega_0^2\alpha^2 \cos^2\theta
-\Omega_0^2\beta^2\sin^4\theta
-2i\alpha\beta \Omega_0kw_0\cos^2\theta\sin\theta\right)^{1/2}\right).
\label{solutions}
\end{equation}
For $\theta=0$, the modes are
\begin{equation}
\sigma_\pm=\frac{2\Omega_0}{1-\epsilon_n}\,(\pm\alpha-i\beta). 
\end{equation}
Imperfect pinning has introduced two low-frequency modes to the system
that are associated with slow vortex motion under the Magnus
force. The modes are underdamped for $\beta<\alpha$, which defines the
regime of low-drag creep that we will explore further.

The general solution with eigenvalue $\sigma_-$ in
eq. (\ref{solutions}) is unstable above a critical wavenumber
$k_c$. To find this wavenumber, we write the solution as
\begin{equation}
\sigma_-=A+iB-\sqrt{C+iD}, 
\end{equation}
where
\begin{equation}
A=\frac{(\alpha-\epsilon_n)}{1-\epsilon_n} kw_0s \qquad
B=-\frac{\Omega_0(1+c^2)}{1-\epsilon_n}\beta \qquad
C=\frac{4\Omega_0^2}{(1-\epsilon_n)^2}(\alpha^2 c^2
-\beta^2 s^4) \qquad
D=-\frac{2\alpha\beta \Omega_0}
{(1-\epsilon_n)^2}kw_0c^2s
\end{equation}
The solution becomes unstable when 
\begin{equation}
{\rm Im}(\sigma_-)=B-{\rm Im}\left(\sqrt{C+iD}\right)=0,
\label{im} 
\end{equation}
where
\begin{equation}
{\rm Im}\left(\sqrt{C+iD}\right)=
(C+D)^{1/4}
\sin\left(\frac{1}{2}\cos^{-1}\left(\frac{C}{\sqrt{C^2+D^2}}\right)\right).
\end{equation}
Combining this equation with eq. (\ref{im}) gives
\begin{equation}
4B^2(B^2+C)=D^2. 
\end{equation}
Solving for $k$ gives the critical
wavenumber $k_c$ above which the system is unstable:
\begin{equation}
k>k_c\equiv2\frac{\Omega_0}{w_0}\frac{(\beta^2+\alpha^2)^{1/2}}
{\alpha}\,
\frac{1+\cos^2\theta}{\sin\theta\cos\theta}, 
\label{kc}
\end{equation}
independent of entrainment. 
The critical wavenumber $k_c$ is minimized for
$\theta=\tan^{-1}(\sqrt{2})$. For $k>>k_c$, we have the approximate
solutions 
\begin{equation}
\sigma_\pm\simeq
\frac{1}{1-\epsilon_n}
((\alpha-\epsilon_n) kw_0\sin\theta\mp
i(\alpha\beta\,\Omega_0kw_0\cos^2\theta\sin\theta)^{1/2})
\qquad\qquad
k>>k_c.
\label{highk}
\end{equation}
For the outer core, $\epsilon_n\simeq 3\times 10^{-3}$ for $x_p=0.05$ 
\citep{ch06}. Though entrainment is essential in producing vortex
pinning, it has a negligible effect on the growth rate of the
instability, and no effect on the critical wavenumber, so we ignore it
in the estimates below. 

The instability arises from coupling between the velocity difference
$\wbf_{np}$ and the neutron vorticity $\omegabf_n$ through the force
of eq. [\ref{df}].  Dissipation damps perturbations for $k<k_c$, but
for $k>k_c$ the finite vortex mobility gives rise to growing
perturbations under the Magnus force.  For $k>>k_c$, the growth rate
scales as $(\alpha\beta w_0)^{1/2}$. For $\beta<<\alpha$, $k_c$ takes
a constant value, but the growth rate of the mode becomes small, going
to zero as $\beta$ goes to zero. In the highly-damped regime,
$\beta>>\alpha$, damping restricts the unstable mode to large $k$,
generally stabilizing the system. There are no unstable modes for
either $\alpha=0$ or $\beta=0$; the instability occurs only if 
vortex motion has components along both $\hat{w}_{np}$ and 
$(\hat{\omega}_n\times\hat{w}_{np})\times\hat{\omega}_n$. 

We now confirm that the viscous stress is negligible for the $\sigma_-$
unstable mode. The
magnetic stress force per unit volume on the charged fluid from
eq.(\ref{paccel}) is 
\begin{equation}
\delta\fbf_m/\rho_p=\frac{\Bbf_{{\rm
eff},0}}{4\pi\rho_p}\cdot\nabla\delta\Bbf_{\rm eff}, 
\end{equation}
while the viscous force is 
\begin{equation}
\fbf_v/\rho_p=\nu_e\nabla^2\vbf_p
\end{equation}
Fourier transforming, and using the induction equation
(\ref{induction}), gives
\begin{equation}
\frac{\delta f_v}{\delta f_m}\sim \frac{\nu_e}{v_B^2}
\left(\frac{B_c}{B_0}\right)^{1/2}\,{\rm Re}(i\sigma_-). 
\end{equation}
This ratio is largest for high wavenumber. In this limit
\begin{equation}
\frac{\delta f_v}{\delta f_m}\sim \Omega_0\frac{\nu_e}{v_B^2}
\left(\frac{B_c}{B_0}\right)^{1/2}
\left(\frac{w_0}{c_T}\right)^{1/2}
(\alpha\beta)^{1/2}
\left(\frac{kc_T}{\Omega_0}\right)^{1/2}
\end{equation}
In the outer core, $\nu_e$ is typically $\sim 10^6$ cm$^2$ s$^{-1}$ 
\citep{cl87,acg05}. 
Below we estimate $\alpha\beta\sim 10^{-18}$.  The hydrodynamic
treatment requires $kc_T<<\Omega_0$, so $\delta f_v/\delta f_m\lap
10^{-9}$ for a typical neutron star with $\Omega_0=100$ \rads\ and
$B_0=10^{12}$ G; the viscous force is negligible compared to the
magnetic force for these low-frequency modes.

\section{Estimates of the instability growth rate}

To obtain the growth rate of the instability, we now estimate the
pinning parameters $\alpha$ and $\beta$ for the vortex creep process.
To make these estimates, we regard the process of vortex creep
as consisting of two distinct states of motion for a given vortex
segment. Most of the time, the vortex segment is pinned. A small
fraction of the time, the vortex segment is translating against a drag
force to a new pinning configuration. The mutual friction force we are
using (eq. \ref{fmf}) is 
\begin{equation}
\fbf/\rho_n=
\omegabf_n\times\wbf_{np}
-\alpha\,\omegabf_n\times\wbf_{np}
+\beta\, \hat{\omega}_n\times(\omegabf_n\times\wbf_{np}). 
\label{fagain}
\end{equation}
This force represents the spatially and temporally averaged force
exerted on the neutron fluid by the creep process. 
The first term is the Magnus force for perfect pinning, while the
remaining terms give the contribution to the force due to vortex
motion. When a vortex segment is unpinned and moving against drag, we take the
force to have the same form, but with different coefficients:
\begin{equation}
\fbf_0/\rho_n=
\omegabf_n\times\wbf_{np}
-\alpha_0\,\omegabf_n\times\wbf_{np}
+\beta_0\, \hat{\omega}_n\times(\omegabf_n\times\wbf_{np}).
\label{fmf0}
\end{equation}
An unpinned vortex segment remains unpinned for a time $t_0\sim d/w_0$,
where $d$ is the distance the segment moves before repinning. This
distance is comparable to the distance between pinning sites
\citep{leb93}. For pinning to flux tubes, 
$d\sim l_\Phi\simeq\sqrt{\Phi_0/B}$, and
the average time that a vortex segment is unpinned is
\begin{equation}
t_0\sim 10^{-14}\,B_{12}^{-1/2}\left(\frac{w_0}{10^6\mbox{
\cms}}\right)^{-1}\mbox{ s }, 
\end{equation}
much shorter than the hydrodynamic timescales of interest. 
Suppose that at any instant, the volume-averaged fraction of vortex
length that is unpinned is $f_v<<1$. 
We now average $\fbf_0$ over 
a volume that contains many vortices, and
over a time long compared to $t_0$ but short compared to hydrodynamic
timescales, to obtain
\begin{equation}
\langle{\fbf_0}/\rho_n\rangle=
\omegabf_n\times\wbf_{np}
-f_v\alpha_0\,\omegabf_n\times\wbf_{np}
+f_v\beta_0\, \hat{\omega}_n\times(\omegabf_n\times\wbf_{np}).
\label{fave}
\end{equation}
Quantities related to the flow are unchanged by the averaging
procedure since the superfluid flow velocity, the vorticity, and the
relative flow velocity are independent of whether
vortices are pinned or not.  The factors of $f_v$ in eq. (\ref{fave})
account for the fact that only the motion of the translating vortex segments
contributes to the mutual friction (see, also,
\citealt{jahanmiri06}). For vortex motion by thermal activation,
$f_v\sim {\rm e}^{-A/T}$, where $A$ is the activation energy for
unpinning and $T$ is the temperature. The value of $f_v$ is
unimportant for the following estimates.

The force of eq. (\ref{fagain}), which is appropriate for vortex
creep, must equal the average force $\langle{\fbf_0}/\rho_n\rangle$,
giving the following relationships: 
\begin{equation}
\alpha=f_v\,\alpha_0
\quad
\mbox{and}
\quad
\beta=f_v\,\beta_0 
\quad
\Rightarrow 
\quad
\frac{\beta}{\alpha}=\frac{\beta_0}{\alpha_0}
\label{alphabeta}
\end{equation}
We now take estimates of $\beta_0/\alpha_0$ to obtain the ratio
$\beta/\alpha$. 
 
The dominant drag process on unpinned vortex segments identified so
far is the scattering of electrons against the magnetized cores of
neutron vortices \citep{als84}. For unpinned vortex segments, vortex
motion can be described in terms of a single drag coefficient, as in
eq. (\ref{drag}), so that 
\begin{equation}
1-\alpha_0=\frac{{\cal R}^2}{1+{\cal R}^2}={\cal R}\,\beta_0, 
\end{equation}
Using recent values of the proton
effective mass,
\citet{sa09} estimate 
\begin{equation}
{\cal R}\simeq 
4\times 10^{-4}-2\times 10^{-5}, 
\end{equation}
giving $\alpha_0\simeq 1$ and $\beta_0\simeq {\cal R}$, so that 
\begin{equation}
\frac{\beta_0}{\alpha_0}=\frac{\beta}{\alpha}\simeq {\cal R}.
\end{equation}
Vortex creep in the outer core is thus well into the low-drag
regime. We stress that, in estimating $\beta_0/\alpha_0$, we have
assumed that the drag relationship given by eq. (\ref{drag}) holds only for
the small fraction $f_v$ of unpinned vortex length; the
volume-averaged quantities $\beta$ and $\alpha$ satisfy
$\beta/\alpha=\beta_0/\alpha_0$, from eq. (\ref{alphabeta}), but there
is no value of ${\cal R}$ that gives both $\beta<<\alpha$ and
$\alpha<<1$, the creep regime of low dissipation in which we are interested.

We now estimate $\beta$. We adopt polar coordinates $(r,\phi,z)$, with
the unperturbed vorticity along $\hat{z}$ and the unperturbed flow
$\wbf_0$ along $\hat{\phi}$, and take the unperturbed flow and vortex
velocity field to be axisymmetric.  In the rotating frame ($\vbf_p=0$), 
the unperturbed vortex velocity from eq. (\ref{vv}) is
\begin{equation}
\vbf_{v,0}=\alpha\,w_0\, \hat{\phi}+\beta\, w_0\,\hat{r} 
=\hat{n}\, v_{v,0}
\end{equation}
where $\hat{n}$ is the average direction of vortex motion. 

For steady spin down of the star, the core superfluid, the charged
components, and the crust are
spinning down at the same rate for a local differential velocity
$w_0$. The creep velocity in this steady state is related to the
spin-down rate by \citep{alpar_etal84a,leb93} 
\begin{equation}
\dot{\Omega}=-2\frac{\Omega}{r}\, \vbf_{v,0}\cdot\hat{r}
=-2\frac{\Omega}{r}\, w_0\,\beta
=\dot{\Omega}_0, 
\end{equation}
where $\Omega$ is the spin rate of the superfluid, $\dot{\Omega}_0$ is
the observed spin down rate of the crust, and $r$ is approximately the
stellar radius $R$. We arrive at the estimate 
\begin{equation}
\beta=\frac{R}{4w_0t_{\rm
age}}
\simeq 10^{-11} \left(\frac{w_0}{10^5\mbox{ \cms}}\right)^{-1}
\left(\frac{t_{\rm age}}{10^4\mbox{ yr}}\right)^{-1}.
\label{ss}
\end{equation}
where $\Omega\simeq\Omega_0$ is assumed.  Eq. (\ref{ss}), in
combination with eq. (\ref{alphabeta}), gives the fiducial values
$\alpha=10^{-7}$ and $\alpha\beta=10^{-18}$. For these values, we
deduce $f_v\sim (\alpha\beta/\alpha_0\beta_0)^{1/2}\sim 10^{-6}$, that
is, most of the vortex length is pinned at any instant.

We can now proceed with estimates of the instability length scale and
growth rate. 
For $\alpha>>\beta$ and $\theta=\tan^{-1}(\sqrt{2})$ in eq. (\ref{kc}),
the critical wavenumber is
\begin{equation}
k_c\simeq 6\,\frac{\Omega}{w_0}=6\times 10^{-3}
\left(\frac{\Omega}{100\mbox{ rad s$^{-1}$}}\right)
\left(\frac{w_0}{10^5\mbox{ \cms}}\right)^{-1}, 
\end{equation}
corresponding to a wavelength $\lambda=2\pi/k\simeq 10$ m. For
$k>>k_c$, the growth rate from eq. (\ref{highk}) is 
\begin{equation}
\frac{1}{2\pi}\,{\rm Im}(\sigma_-)\simeq 2\, 
\left(\frac{\alpha\beta}{10^{-18}}\right)^{1/2}
\left(\frac{\Omega}{100\mbox{ rad s$^{-1}$}}\right)^{1/2}
\left(\frac{w_0}{10^5\mbox{ \cms}}\right)^{1/2}
\left(\frac{\lambda}{1\mbox{ m}}\right)^{-1/2}
\mbox{ yr$^{-1}$}
\end{equation}

The hydrodynamic
treatment is restricted to $c_Tk<<\Omega$. 
To estimate how high the growth rate could be,
we consider a maximum wavenumber defined by 
$c_Tk_{\rm max}=0.1\,\Omega$, where 
$c_T\simeq 10^{-1}\, (\Omega/100\mbox{ \rads})^{1/2}$ \cms. 
The growth rate at this wavenumber, from eq. (\ref{highk}), is 
\begin{equation}
\frac{1}{2\pi}\,{\rm Im}[\sigma_-(k_{\rm max})]\simeq 0.2 
\left(\frac{\alpha\beta}{10^{-18}}\right)^{1/2}
\left(\frac{\Omega}{100\mbox{ rad s$^{-1}$}}\right)^{3/4}
\left(\frac{w_0}{10^5\mbox{ \cms}}\right)^{1/2}
\mbox{ d$^{-1}$}. 
\label{highsigma}
\end{equation}
For $\Omega=100$ rad s$^{-1}$, the corresponding wavenumber is
$k\simeq 100$ cm$^{-1}$.  Eq. (\ref{highsigma}) does not
represent a physical limit, but only the restrictions of the
hydrodynamic treatment; the instability could continue to exist also
for wavenumbers in the regime $kc_T>\Omega$. If vortex creep is in
the strongly-damped regime $\beta>>\alpha$, contrary to the estimates
here, there is still a broad window for instability. Requiring
$k_c<k_{\rm max}$ gives
\begin{equation}
\beta<2\times 10^4\, \left(\frac{w_0}{10^5\mbox{ \cms}}\right)
\left(\frac{\Omega}{100\mbox{ rad s$^{-1}$}}\right)^{-1/2}\, \alpha, 
\end{equation}
and the star will be unstable at some wavenumber that is consistent
with the hydrodynamic approximation $c_Tk<<\Omega$. 

\section{Timing noise driven by superfluid turbulence?}  

The superfluid flow could quickly become turbulent through the
instability identified here. Turbulent flow would exert random torques
on the crust, possible contributing to observed timing anomalies such
as timing noise and glitches. We now estimate the spectrum of turbulent torque
fluctuations using the similarity argument of \citet{kolmogorov41}
and \citet{obukhov41}. In this argument, the properties of the
turbulent spectrum follow from the assumption that turbulent energy is
being dissipated at a rate $\dot{\epsilon}$ per unit mass at some high
wavenumber. 

The torque fluctuation per unit mass of fluid for a mode of wavenumber $k$ is 
\begin{equation}
\delta N(k)\,dk\simeq\pm l\,(\Delta v) f
\end{equation}
where $l\sim 1/k$ is the characteristic dimension of a turbulent cell,
and $\Delta v$ is the characteristic change in flow speed over a time
$1/f$, and $fl\sim \Delta v$. 
The units of $\delta N(k)$ are cm$^3$ s$^{-2}$, and the only possible
combination of $\dot{\epsilon}$ and $k$ with the correct units is 
\begin{equation}
\delta N(k)=\pm A (\dot{\epsilon})^{2/3}\,k^{-5/3}, 
\end{equation}
where $A$ is a dimensionless constant. 
The time average of $\delta N(k)$ is zero. 

Integrated over the star of radius $R$, the contributions to the total
torque by different cells of wavenumber $k$ add incoherently, and the 
fluctuation in the total torque at wavenumber $k$ scales as
\begin{equation}
\delta N_{\rm tot}(k)\propto \sqrt{M}\,\delta N(k),
\end{equation}
where $M$ is the characteristic number of turbulent cells of
wavenumber $k$, 
\begin{equation}
M=\left(\frac{R}{l}\right)^d. 
\end{equation}
Here $d$ is the dimensionality of the turbulence. Isotropic turbulence
corresponds to $d=3$, while highly-polarized turbulence corresponds to
$d=1$. The turbulence should be polarized to some extent by rotation
(see, \eg, \citealt{andersson_etal07}).

The fluctuations in the total torque scale as
\begin{equation}
\delta N_{\rm tot}(k)\propto\pm k^{d/2}k^{-5/3}, 
\end{equation}
and the spectral power of torque fluctuations (spectral index
$\alpha_t$) is 
\begin{equation}
\frac{d\vert\delta N_{\rm tot}(f)\vert^2}{df}
\propto f^{-13/3+d}\equiv f^{\alpha_t}\longrightarrow f^{-3.3} 
\mbox{ to } f^{-1.3}.  
\end{equation}
Let the rotation phase of a pulsar be $\phi(t)$, so that $d\phi(t)/dt$
is the spin rate. 
A torque variation $\delta N_{\rm tot}(t)$ produces a variation in
the rotational phase $\delta\phi(t)$, where 
\begin{equation}
\delta N_{\rm tot}(t)=
2\pi I_c\, \frac{d^2\delta\phi}{dt^2}, 
\end{equation}
 and $I_c$ is the moment of inertia of the
crust plus any other components of the star tightly coupled to the
crust.  The spectrum of phase residuals $\delta\phi$
(spectral index $\alpha_p$) is therefore
\begin{equation}
\frac{d\vert\delta\phi(f)\vert^2}{df}\propto f^{-25/3+d}
\equiv f^{\alpha_p}
\longrightarrow f^{-7.3} \mbox{ to } f^{-5.3}.  
\end{equation}
``Torque noise'' corresponds to $\alpha_p=-5$. We conclude that
turbulence can produce such a power spectrum only if the turbulence
is nearly isotropic ($d=3$). From the scaling argument given here,
turbulence cannot produce ``frequency noise'' ($\alpha_p=-3$) nor
``phase noise'' ($\alpha_p=-1$). 
In early work on 45 pulsars,
\citet{dalessandro_etal95} reported power spectra consistent with a
range $-1\le \alpha_p<-5$. More recent estimates are given in Table 1
for six radio pulsars and two magnetars.  These results should be
regarded as illustrative only; the power spectra of many pulsars do
not follow simple power laws \citep{cd85}, and some sources show
quasi-oscillatory behavior; see \citet{hobbs_etal10} for a discussion
of these issues.  Though timing noise is a complicated process that is
difficult to characterize, we note that the basic power spectra
predicted by turbulent torque fluctuations are generally consistent
with observed power spectra. 

\begin{table}
\begin{tabular}{lll}
\hline
SOURCE & SPECTRAL INDEX  & REFERENCE \\
\hline 
PSR 0823+26 & $\alpha_t=-1.16\pm 0.48$ & \citet{baykal_etal99} \\
            & $\alpha_t=-0.39\pm 0.36$  & \\
PSR 2021+51    & $\alpha_t=-1.95\pm 0.74$ & \citet{baykal_etal99} \\
               & $ \alpha_t=-1.64\pm 1.40$ & \\
PSR 1706-16 & $\alpha_t=-2.41\pm 0.79$ & \citet{baykal_etal99} \\
            & $\alpha_t=-1.94\pm 1.33$ & \\
PSR 1749-28 & $\alpha_t=-0.88\pm 0.05$ &  \citet{baykal_etal99} \\
            & $\alpha_t=-1.05\pm 0.83$ & \\
PSR B1509-58 & $\alpha_p=-4.6\pm 1.0$ & \citet{livingstone_etal05} \\
PSR B0628-28 & $\alpha_p\simeq -1$ ($t<2$ yr) & \citet{hobbs_etal10} \\
PSR B0628-28 & $\alpha_p\simeq -5$ ($t>2$ yr) & \citet{hobbs_etal10} \\
SGR 1900+14 & $\alpha_t=-3.7\pm 0.6$ & \citet{woods_etal02} \\
SGR 1806-20 & $\alpha_t=-3.6\pm 0.7$ & \citet{woods_etal02} \\
\hline

\end{tabular}
\caption{Estimates of spectral indices. The second entry for PSRs
0823+26, 2021+51, 1706-16, and 1759-29 follows from a fit to the arrival
times to a higher-order polynomial; see \citealt{baykal_etal99} for
details. PSR B0628-28 shows $\alpha_p\simeq -1$ over timescales of less
than two years, and $\alpha_p\simeq -5$ over longer timescales.}
\end{table}

\section{Discussion and conclusions}

We have identified a dissipation-driven instability that could operate
in the fluid of the neutron star core over length scales shorter than
$\sim 10$ m and over timescales as fast as days.  The instability is a
result of the forcing the vortices against the rigid array of flux
tubes.  The superfluid flow of the in the outer core could become
turbulent if dissipative processes do not suppress the instability.  The
excitation of Kelvin waves vortices \citep{link04} will lead to
dissipation, but is probably not nearly strong enough to put vortex
creep into the highly-dissipative regime that would be required to quench
the instability; this process damps rotational perturbations over many
spin periods, and so is in the low-drag regime. Other dissipation
mechanisms of possible relevance require study. If the core fluid
becomes turbulent, the closest experimental analogue might be grid
turbulence that has been well studied in superfluid helium
\citep{smith_etal93}. In our previous work on the crust
\citep{link12a}, we found that the flow is unstable there as well. We
conclude that superfluid flow throughout much of a neutron star is
forced unstable by the pinning stresses on the fluid that arise as the
star spins down. 

To illustrate the basic instability, we have taken a simplified mutual
friction force in which the pinning coefficients $\alpha$ and $\beta$
are constants. The estimates of the instability growth rate given
here should be regarded as rough estimates. For thermally-activated
vortex creep, the pinning coefficients will have exponential dependence on
the scalar $w_{np}$
\citep{alpar_etal84a,leb93}. We expect that this strong velocity
dependence will significantly enhance the growth rate of the
instability, without significantly changing the critical wavenumber so
long as vortex motion is in the regime of low drag. Further work is
needed on this issue. 

Turbulent flow could create torque fluctuations with spectra that are
consistent with those measured in many radio pulsars and in
magnetars. While the simple scaling argument given in \S 5 suggests
that turbulent torques cannot produce ``phase noise'' or ``frequency
noise'', it would be interesting to further explore the possibility
that steeper power spectra result from turbulent torques. In future
work we will determine if the instability identified in this
paper leads to fully-developed turbulence, and estimate the
magnitude of the fluctuations. 


\section*{Acknowledgments}
We thank the anonymous referee for particularly useful constructive
criticism. 


\bibliography{references}

\label{lastpage}

\end{document}